# Compact Modeling of 0.35 μm SOI CMOS Technology Node for 4 K DC Operation using Verilog-A

A. Akturk[†], K. Eng[‡], J. Hamlet[‡], S. Potbhare[†], E. Longoria[‡], R. Young[‡], M. Peckerar[†], T. Gurrieri[‡], M. S. Carroll[‡], N. Goldsman[†]

[†]Department of Electrical and Computer Engineering
University of Maryland College Park, MD, 20742
[‡]Sandia National Laboratories, Albuquerque, NM, 87185
akturka@umd.edu

## Abstract

Compact modeling of MOSFETs from a 0.35 μm SOI technology node operating at 4 K is presented. The Verilog-A language is used to modify device equations for BSIM models and more accurately reproduce measured DC behavior, which is not possible with the standard BSIM model set. The Verilog-A approach also allows the embedding of nonlinear length, width and bias effects into BSIM calculated curves beyond those that can be achieved by the use of different BSIM parameter sets. Nonlinear dependences are necessary to capture effects particular to 4 K behavior, such as current kinks. The 4 K DC behavior is reproduced well by the compact model and the model seamlessly evolves during simulation of circuits and systems as the simulator encounters SOI MOSFETs with different lengths and widths. The incorporation of various length/width and bias dependent effects into one Verilog-A / BSIM4 library, therefore, produces one model for all sets of devices for this technology node.

## I. INTRODUCTION

Low temperature operation of metal oxide semiconductor field effect transistor (MOSFET) based electronic components has become a field of increasing interest over the last decade due to the fast growing pace of the satellite industry and space exploration [1,2], as well as the emergence of high performance computing applications [3-5]. Complementary MOS (CMOS) integrated circuits generally show improved performance upon cooling due to increased drive currents in saturation and linear regions of operation [6]. The reduction in temperature results in both changes in the classic behavior (e.g., higher mobility [6,7], reduced sub-threshold slope [8] and shifts in threshold voltage [7,9]) as well as the emergence of new effects especially in non-optimized structures for cryogenic operation (e.g., possible hysteresis [10,11], kink effects [10,12] and hot-carrier lifetime degradation [13]). The additional nonlinearities observed in current-voltage curves measured at extremely low temperatures give rise to challenges in the design of reliable and predictable circuits and systems. Typically, CMOS processes are designed for 300 K operation, and very low temperatures introduce uncertainty into the performance and long-term reliability of low temperature electronics because design tools and methodologies are built for room temperature and not for cryogenic temperatures in the range of 4 K or less.

In this paper, we demonstrate that closed-form expressions relating critical component parameters (thresholds, mobilities, effective channel lengths, etc.) can be inserted in SPICE simulators using the Verilog-A behavioral modeling language [14,15]. We show that these compact models provide accurate representations of the measured DC component performance down to 4 K, and handle the introduction of new physical processes such as low-temperature current kink. The ability to model low temperature device operation is critical for the design of predictable, reliable and optimized circuits, as well as achieving the time and investment savings on experiments and prototype development.

## II. MOSFET Compact Modeling at 4 K

Closed form equations for critical parameters, "compact models", suitable for incorporation into generic circuit simulators (generally referred to as SPICE simulators) are commonly used for 300 K. BSIM models, which are a collection of compact analytical equations, are industry-standards for simulating electrical performance of state-of-the-art devices and integrated circuits [16,17]. These models are widely employed by the semiconductor industry due to the speed and accuracy of BSIM-calculated currents and voltages, which in turn give rise to first-pass designs saving resources and time. However, BSIM models are developed for device operation at room temperature and are generally applicable to devices operating ±100 K of this temperature, but can fail to accurately capture the physics of 4 K MOSFET behavior.

Compact model parameters are often obtained from extraction programs like IC-CAP [18,19] using measured MOSFETs over a wide range of lengths and widths. It was found that the MOSFET behavior at 4 K deviates from room temperature enough, however, that IC-CAP parameter extraction for the BSIM-SOI models does not provide good fits to the observed data, Fig 1. Poor simulation of experimentally measured behavior occurs in many regions of operation, especially the subthreshold and in higher source-drain bias relative to the gate that leads to kink behavior [10,12,20]. Modification of the BSIM equations is required to capture the physics behind the cryogenic behavior that is not available in the room temperature BSIM model.

Many industry-standard device simulators such as Cadence's Spectre [21] or Silvaco's SmartSpice [22] have BSIM equations hardcoded into their proprietary source code. Even though the BSIM model equations are open-source, the development of a custom-SPICE simulator to incorporate cryogenic effects into BSIM is likely to have a limited use since its incorporation into industry-standard commercial simulators would require the coding of new program lines into these proprietary simulators, and also might require the approval by the Compact Model Council [23].

To overcome these limitations, we use the analog behavioral programming language Verilog-A that offers flexibility for modifying equations, integrability to modern commercial device simulators, as well as compatibility with other device and passive models for simulating part of a circuit in Verilog-A [14,15]. Verilog-A can be used to tailor BSIM equation and parameter lists that can better address the 4 K behavior. This enables us to develop 4 K MOSFET individual device and circuit simulation formalisms that are adaptable to efficient industry-standard modeling tools such as Cadence's Spectre, Silvaco's SmartSpice and Agilent's ADS [24].

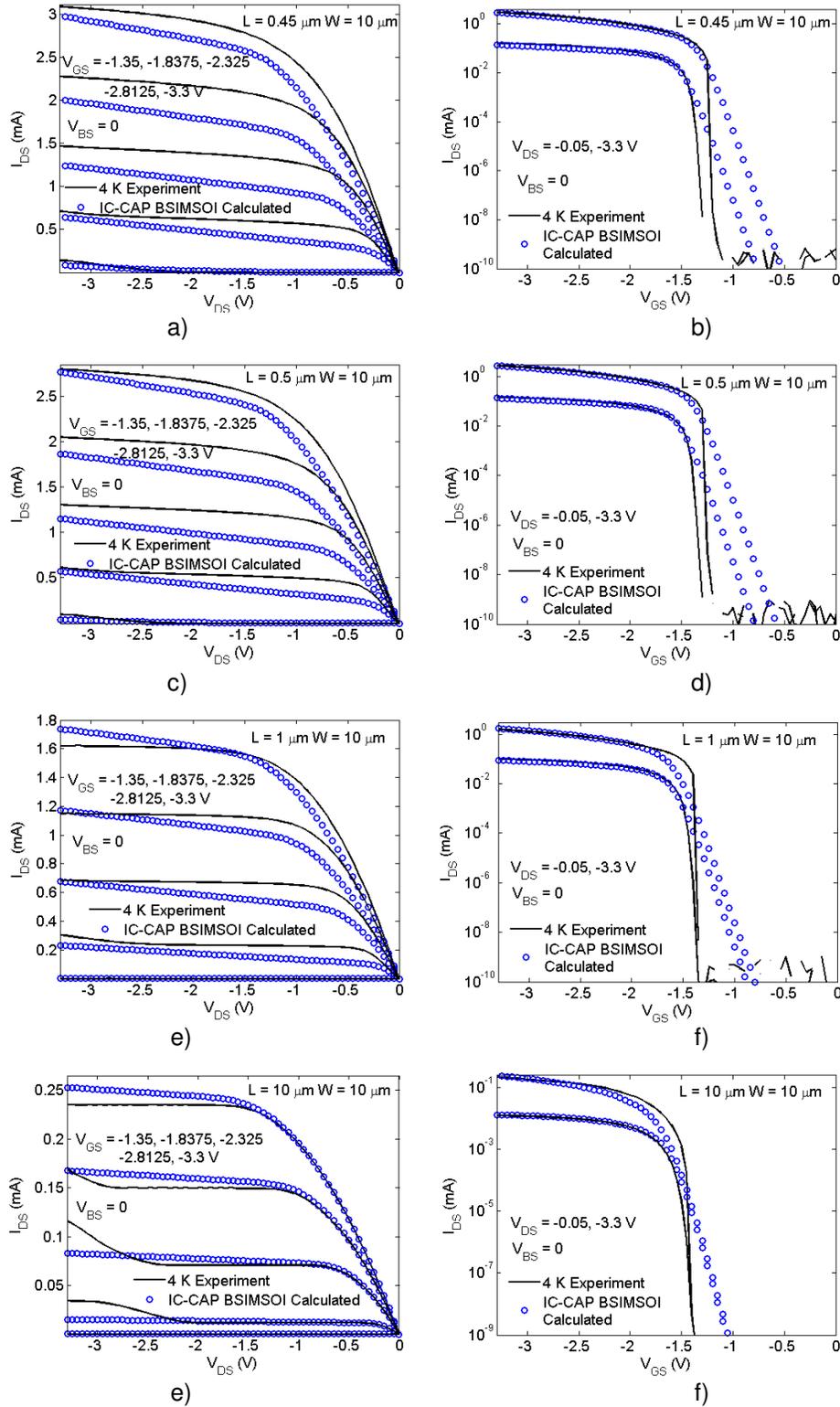

**Figure 1**: Measured (solid lines) and IC-CAP simulated (circles) drain-to-source current versus drain-to-source and gate-to-source voltage curves of different length (a-b: L = 0.45 µm, c-d: L = 0.5 µm, e-f: L = 1.0 µm, g-h: L = 10.0 µm, and W = 10.0 µm for a-f) p-channel SOI MOSFETs.

### III. 4 K CMOS7 SOI-MOSFET Modeling

Current-voltage measurements of *p*- and *n*-channel partially-depleted silicon-on-insulator (SOI) MOSFETs that were fabricated in the Sandia National Laboratories 0.35 µm CMOS7 process line [25] were taken at 4 K using a Lakeshore cryogenic probe station. The sample was thermally mounted to a copper chuck inside an evacuated test chamber, and then cooled by continuous-transfer of liquid Helium. Two-terminal DC measurements of the source-drain current, $I_{SD}$, were later recorded as a function of source-drain ($V_{SD}$), -gate ($V_{SG}$) and, -body ($V_{SB}$) biases using an Agilent B1500 semiconductor device analyzer, while keeping the temperature fixed at 4 K.

We measured various length and width CMOS7 MOSFETs to cover the available design space. More specifically, for the model development we used 0.35, 1.0 and 10.0 µm long *n*-channel, as well as 0.45, 0.5, 1.0 and 10.0 µm long *p*-channel SOI MOSFETs with widths ranging from 0.8 µm to 10.0 µm, and applied biases between two terminals no larger than 3.5 V in magnitude.

To model these *n*- and *p*-channel partially depleted CMOS7 SOI MOSFETs, we followed the algorithm shown in Fig. 2. We first obtained a BSIM4 model translated into Verilog-A [26]. We also achieved a capability to import and simulate this Verilog-A / BSIM4 in commercial compact model simulators. We note that the parameters were extracted taking the nominal temperature, TNOM in BSIM4, and the global simulator temperature, $temperature in the device simulator, as both equal to room temperature 300 K. Example extracted model parameters that provided close fits to L = 1 µm and W = 10 µm *p*-channel SOI MOSFET are indicated in Table 1.

**Table 1**. Example model parameters initially extracted for BSIM4

| **Parameter** | **Value** |
|---|---|
| VTH0 (Threshold voltage) | -1.345 V |
| U0 (Low-field mobility) | 544.4 cm$^2$/Vs |
| RDSW (Zero bias LDD resistance per unit width) | $2 \times 10^3$ Ω.µm$^{WR}$ where channel width dependence parameter WR = 1.06 |
| Vsat (Saturation velocity) | $8.265 \times 10^4$ m/s. |

The calculated fit to this *p*-channel SOI MOSFET partially corroborated with the measured curves in the linear and saturation regions of operation, however, it showed a much higher subthreshold slope (due to $temperature being equal to 300 K) and did not capture the 4 K current kinks. Additional equations were incorporated into the Verilog-A / BSIM 4 model to improve the agreement between simulation and experiment in all regions of device operation. This was mainly achieved using transfer functions at the gate input terminal.

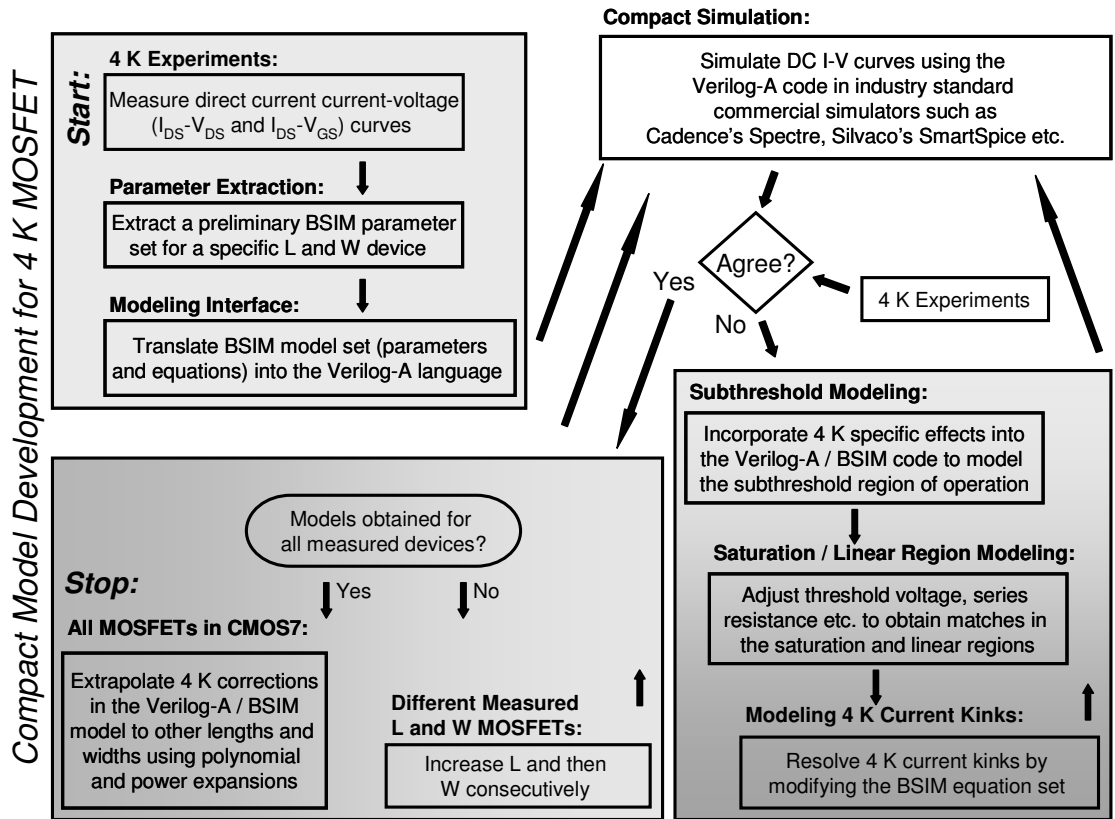

**Figure 2**: Algorithm Flowchart

The first transfer function manipulated the actual gate bias magnitude, Vgs, seen by the device. This produced lower subthreshold slopes. This initial correction which enhanced the current above threshold and suppressed the subthreshold current was achieved using a fourth-order polynomial expansion:

$$Vgs^{eff,1} = Vgs - f_1(Vgs) \text{ and} \tag{1}$$

$$f_1(Vgs) = p4 \times Vgs^4 + p3 \times Vgs^3 + p2 \times Vgs^2 + p1 \times Vgs + p0, \tag{2}$$

where $Vgs^{eff,1}$ indicated the new effective gate potential after the first correction, and expansion coefficients p4, p3, p2, p1, p0 were set to 0.5081, -5.212, 19.64, -32.32 and 19.62, respectively, for the L = 1 µm and W = 10 µm *p*-channel SOI MOSFET.

Here, the calculated correction to the subthreshold slope is less than that obtained using the standard subthreshold slope formula (i.e., a linear change with zero offset). The measured subthreshold slope is significantly shallower than what would be predicted by the standard subthreshold dependence on temperature. Decreasing subthreshold slope with increasing drain bias is furthermore observed. The physical mechanisms responsible for this are under investigation to understand the relative importance of factors such as heating and changing channel dopant ionization near the drain-channel junction due to impact ionization related effects.

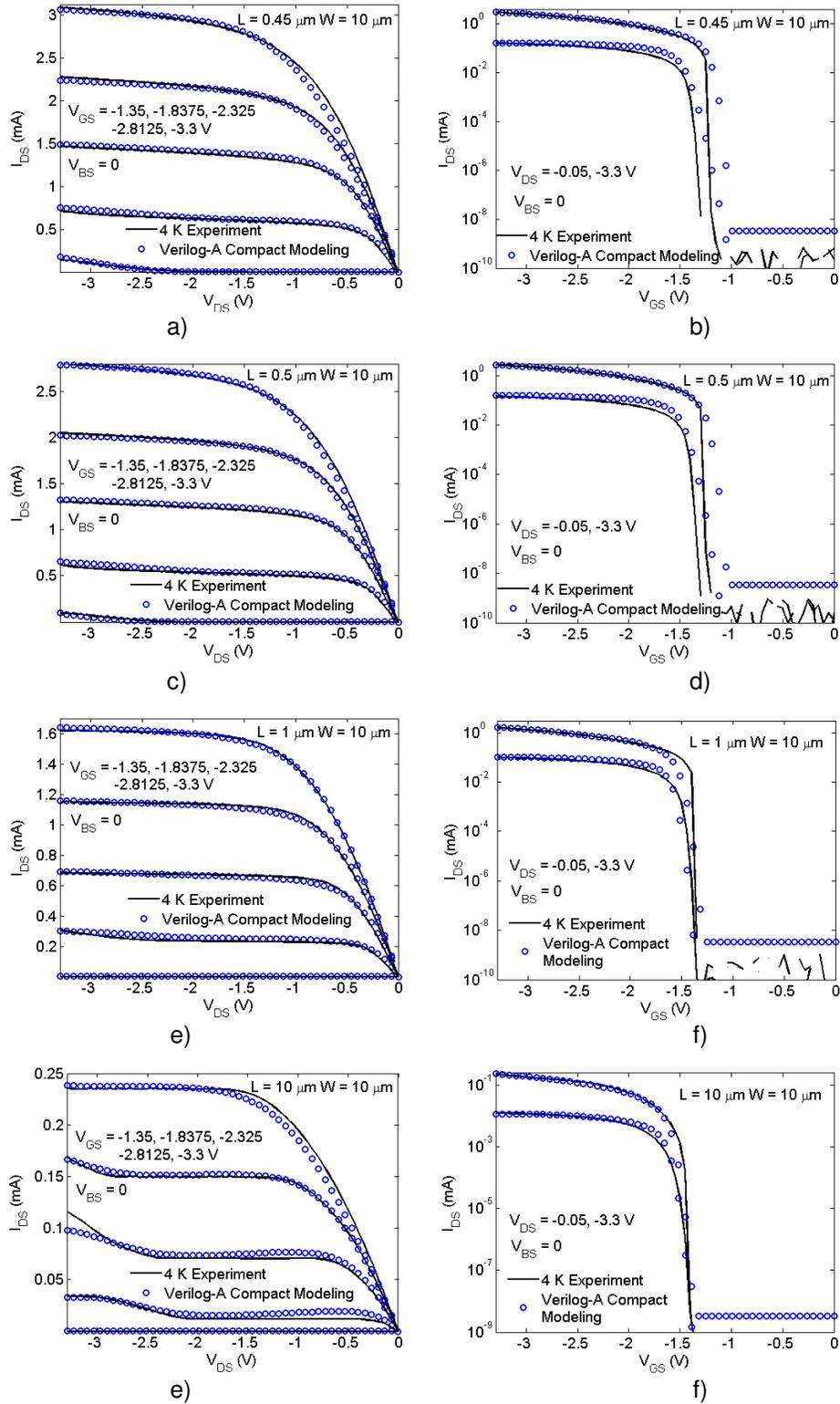

**Figure 3**: Measured and simulated drain-to-source current versus drain-to-source and gate-to-source voltage curves of different length *p*-channel SOI MOSFETs. Simulation results are calculated using the proposed compact Verilog-A / BSIM modeling in Cadence's Spectre.

Compact modeling of the current kinks at 4 K represents an additional challenge because of the significantly nonlinear bias dependence. Current kinking has been attributed to band-to-band impact ionization affecting the dopant ionization level in the channel and substrate [10-12]. The kinks appear at high drain-channel fields peaking at low gate and high drain biases. As the gate bias increases, the associated drain bias needed for the onset of a kink rises in magnitude, suggesting a threshold drain junction field for the start of a current kink.

The associated correction to the gate terminal bias to reproduce current kinks was therefore taken as a function of drain voltage as well as gate bias. For example, for L = 1 µm and W = 10 µm *p*-channel SOI MOSFET the combined effect could be taken as a product of two functions that depended on either gate or drain bias. The gate bias dependency was chosen as: $f_2(Vgs) \times g(Vgs)$, where $f_2(Vgs)$ was a fifth-order polynomial that was approximately constant above threshold and quickly dropped below threshold to compensate for the increase in $g(Vgs)$ in that gate bias range. $g(Vgs)$ was an exponential function of the gate bias set to exp[ -(|Vgs|-|VTH0|) / alpha]. Alpha was a constant that reflected the vanishing effect of current kinks at high gate biases and their importance around the threshold voltage for the *p*-channel devices with lengths less than 1 µm. The drain bias dependency of the second correction to the gate, furthermore, was expressed using $h_1(Vds)$, which was a rational function with a second order numerator and a second order denominator. This functional form was chosen to act like a high-pass filter with amplification by enhancing the gate bias at high drain biases and suppressing the correction at low. Thus the overall gate bias including the corrections to the voltage seen from the gate terminal became:

$$Vgs^{eff} = Vgs - f_1(Vgs) + f_2(Vgs) \times g(Vgs) \times h_1(Vds) \qquad (3)$$

The saturation value of the drain bias, Vdsat, was also adjusted to better estimate the device current in the linear region of operation. Specifically, the Vdsat was set to ×0.6 of its value calculated using the standard BSIM4 equations for the L = 1 µm and W = 10 µm *p*-channel SOI MOSFET. A device length dependency was coded to Vdsat in addition to changing the drain resistance as a function of length. This approach was taken since the slopes of the current curves in the linear region were a strong function of length, which was not as well modelled through incorporation of additional series drain resistance.

Figures 3a-3f show the calculated (by Cadence's Spectre) and measured drain-to-source current versus drain-to-source (saturation and linear regions) voltage as well as gate-to-source (mainly the subthreshold region) voltage curves of *p*-channel CMOS7 SOI MOSFETs with W = 10 µm and channel lengths of 0.45 (Figs. 3a-3b), 0.5 (Figs. 3c-3d) and 1 (Figs. 3e-3f) µm. Reasonable matches between experimental and simulated data are obtained. For this channel length range, there is a subtle onset of current kink at low gate and high drain biases. We also note that the measured subthreshold slope decreases with increasing drain bias. The present form of the corrections results in a lower slope with higher drain bias as well; however, the calculated slope is slightly higher than measured. To obtain better agreement in the subthreshold region of operation, in some instances, the calculated BSIM subthreshold slope, n, is further modified to be a function of drain-to-source bias through multiplication by a continuous function of Vds that is rapidly changing at small drain biases and gives a small constant value at high drain biases. The calculated and measured currents agree well in a large range of bias values and channel lengths and are considerably better fits compared to those in Fig. 1.

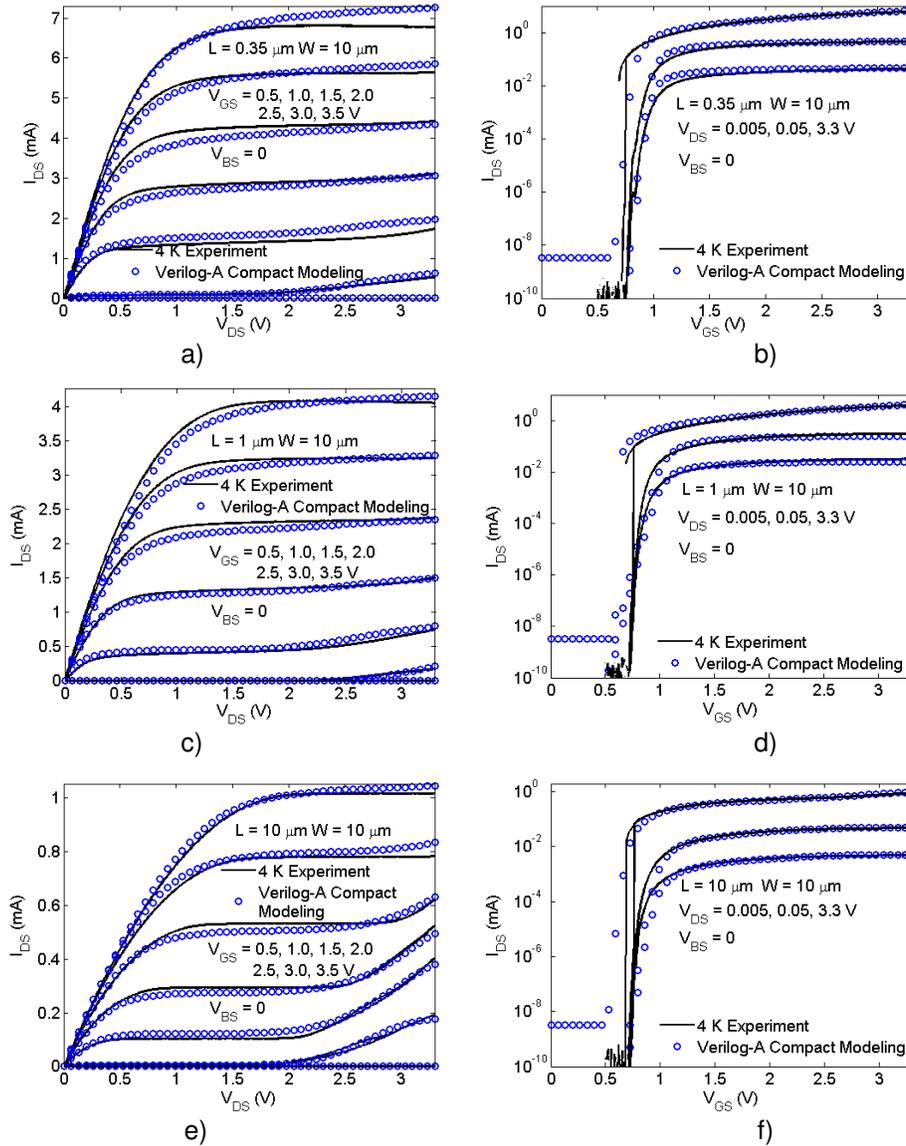

**Figure 4**: Measured and simulated drain-to-source current versus drain-to-source and gate-to-source voltage curves of different length *n*-channel SOI MOSFETs. Simulation results are calculated using the proposed compact Verilog-A / BSIM modeling in Cadence's Spectre.

To incorporate the effects of different channel lengths into current-voltage calculations, we empirically calculated the expansion coefficients of the aforementioned polynomials for the shown lengths, and extrapolated these coefficients for lengths ranging from 0.45 μm to 1 μm. Additionally, we took the previously mentioned mobility parameter, U0, as a linear function of channel length for *p*-SOI MOSFETs while keeping other parameters fixed for all lengths.

For the longest channel *p*-SOI MOSFET with measured curves shown in Figs. 3g-3h, the current kink was more pronounced. As the channel length increased, we saw a consistent increase in the strength and appearance of current kinks. To incorporate this effect, in addition to what had already been considered for the shorter length devices, we needed to modify the second correction to the gate terminal bias mentioned before. We removed the exponential decay embedded in $g(Vgs)$, and replaced the correction $f_2(Vgs) \times g(Vgs) \times h_1(Vds)$ with a convoluted function of drain and gate biases: $h_2(f_3(Vgs),Vds)$ where $h_2$ was a rational function with a second order numerator and a second order denominator, and $f_3(Vgs)$ was similar to $f_2(Vgs)$ in shorter length devices.

In Figs. 3g-3h, we compare our calculated results for the L = 10 µm and W = 10 µm *p*-SOI MOSFET with those measured, and the comparison shows reasonable agreement. Furthermore, we note that we use a single Verilog-A / BSIM4 library to simulate all *p*-channel devices. Minimizing changes to the manual library is desirable to alleviate the burden on the circuit designer and reduce errors.

Modeling of *n*-channel devices was similar to that of *p*-channels. The *n*-channel currents in the saturation region of operation were higher than *p*-channel, as expected, however percentage-wise the higher drive of *n*-channel devices compared to *p*-channels decreased with decreasing channel lengths. Specifically, long channel *n*-SOI MOSFETs supplied approximately ×4 higher current than long channel *p*-SOI MOSFETs, where in contrast, as the device length lowered the drive reduced to less than ×2. Furthermore, in the saturation region of operation the shorter channel *n*-SOI MOSFETs showed negative differential resistances which indicated self-heating and importance of electron-phonon interactions on channel mobility. The *n*-channel devices, furthermore, showed slightly more pronounced current kinks compared to *p*-channel devices, which might be due to their higher current drives, and different impact ionization coefficients as well as substrate dopant ionizations.

To factor effects specific to *n*-channel devices into the Verilog-A / BSIM4 library, similar functional forms were developed as those used in characterization of *p*-channel devices. The only difference in functional dependence between the *n*- and *p*-channel devices was the additional length dependency in some electrical parameters. For the *n*-SOI MOSFETs, length effects were also incorporated into threshold voltage, VTH0, zero bias LDD resistance per unit width, RDSW, and saturation velocity, Vsat, as well as low field mobility, U0. The Verilog-A / BSIM4 curves calculated by Cadence's Spectre simulator for the *n*-channel SOI MOSFETs are compared with those experimentally measured at 4 K in Fig. 4. Reasonable agreement between the two-sets of curves is obtained, which is again significantly better than the simpler fitting approach shown in Fig. 1.

## IV. CONCLUSION

We describe compact modeling of MOSFETs from a 0.35 µm SOI technology node at 4 K that uses the Verilog-A language to modify device equations for BSIM models and more accurately reproduce measured DC behavior, which is not possible with the standard BSIM model set. The Verilog-A approach also allows us to embed nonlinear length, width and bias effects into BSIM calculated curves beyond those that can be achieved by the use of different BSIM parameter sets.

Effects particular to 4 K behavior, such as current kinks, are inserted into the calculated current-voltage curves, in addition to standard classical changes in the temperature dependent MOSFET operation. This is partially achieved by modifying the gate bias seen by the device, and manipulating it through the use of polynomials and rational functions to compensate for the change in subthreshold slope, deviations from calculated currents, and observed current kinks. We further incorporated the length effects into the compact 4 K CMOS7 SOI MOSFET model by empirically calculating sets of expansion coefficients for the aforementioned corrective functions for different channel lengths. The model seamlessly evolves during simulation of circuits and systems as the simulator encounters SOI MOSFETs with different lengths and widths. The incorporation of various length/width and bias dependent effects into one Verilog-A / BSIM4 library, therefore, produces one model for all sets of devices for this technology node.